\documentclass{elsart}

\usepackage{harvard}
\usepackage{epsfig}

\usepackage{amssymb}

\def\url#1{{\ttfamily\def\/{/\discretionary{}{}{}}#1}}

\def\msun{\mbox{${\rm M}_{\odot}$}}
\def\etal{{\it et al.}}

\begin{document}

\begin{frontmatter}
\title{Gasoline: An adaptable implementation of TreeSPH}
\author[mac]{J.W. Wadsley}
\author[uz]{J. Stadel}
\author[uw]{T. Quinn}
\address[mac]{Department of Physics and Astronomy, McMaster University, Hamilton, Canada\thanksref{email}}
\address[uz]{Institute for Theoretical Physics, University of Zurich, Switzerland} 
\address[uw]{Astronomy Department, University of Washington, Seattle, Washington, USA}

\thanks[email]{E-mail: wadsley@mcmaster.ca}

\begin{abstract}
The key algorithms and features of the Gasoline code for parallel
hydrodynamics with self-gravity are described.  Gasoline is an
extension of the efficient Pkdgrav parallel $N$-body code using
smoothed particle hydrodynamics.  Accuracy measurements, performance
analysis and tests of the code are presented.  Recent successful
Gasoline applications are summarized.  These cover a diverse set of
areas in astrophysics including galaxy clusters, galaxy formation and
gas-giant planets.  Future directions for gasdynamical simulations in
astrophysics and code development strategies for tackling cutting edge
problems are discussed.
\end{abstract}

\end{frontmatter}

\section{Introduction}
\label{intro}

Astrophysicists have always been keen to exploit technology to better
understand the universe.  $N$-body simulations predate the use of
digital computers with the use of light bulbs and light intensity
measurements as an analog of gravity for manual simulations of a
many-body self-gravitating system \cite{holmberg}.  Astrophysical
objects including planets, individual stars, interstellar clouds, star
clusters, galaxies, accretion disks, clusters of galaxies through to
large scale structure have all the been the subject of numerical
investigations.  The most challenging extreme is probably the
evolution of space-time itself in computational general relativistic
simulations of colliding neutron stars and black holes.  Since the
advent of digital computers, improvements in storage and processing
power have dramatically increased the scale of achievable simulations.
This, in turn, has driven remarkable progress in algorithm
development.  Increasing problem sizes have forced simulators who were
once content with $\mathcal{O}(N^2)$ algorithms to pursue more complex
$\mathcal{O}(N\,log\,N)$ and, with limitations, even $\mathcal{O}(N)$
algorithms and adaptivity in space and time.  In this context we
present Gasoline, a parallel $N$-body and gasdynamics code, which has
enabled new light to be shed on a range of complex astrophysical
systems.  The discussion is presented in the context of future
directions in numerical simulations in astrophysics, including
fundamental limitations in serial and parallel.

We begin by tracing past progress in computational astrophysics with
an initial focus on self-gravitating systems ($N$-body dynamics) in
section~\ref{gravity}.  Gasoline evolved from the Pkdgrav parallel
$N$-body tree code designed by \cite{stadel}.  The
initial modular design of Pkdgrav and a collaborative programming
model using CVS for source code management has facilitated several
simultaneous developments from the Pkdgrav code base.  These include
inelastic collisions \citeaffixed{DCR00}{e.g. planetesimal dynamics,}, gas
dynamics (Gasoline) and star formation.  In section~\ref{grav} we
summarize the essential gravity code design, including the parallel
data structures and the details of the tree code as applied to
calculating gravitational forces.  We complete the section with a
brief examination of the gravitational force accuracy.

In section~\ref{gas} we examine aspects of hydrodynamics in
astrophysical systems to motivate Smoothed Particle Hydrodynamics
(SPH) as our choice of fluid dynamics method.  We describe the
Gasoline SPH implementation in section~\ref{SPH}, including neighbour
finding algorithms and the cooling implementation.  

Interesting astrophysical systems usually exhibit a large range of
time scales.  Tree codes are very adaptable in space; however,
time-adaptivity has become important for leading edge numerical
simulations.  In section~\ref{stepping} we describe our hierarchical
timestepping scheme.  Following on in section~\ref{performance} we
examine the performance of Gasoline when applied to challenging
numerical simulations of real astrophysical systems.  In particular,
we examine the current and potential benefits of multiple timesteps
for time adaptivity in areas such as galaxy and planet formation.  We
present astrophysically oriented tests used to validate Gasoline in
section~\ref{tests}.  We conclude by summarizing current and proposed
applications for Gasoline.

\section{Gravity}\label{gravity}

Gravity is the key driving force in most astrophysical systems.  With
assumptions of axisymmetry or perturbative approaches an impressive
amount of progress has been made with analytical methods, particularly
in the areas of solar system dynamics, stability of disks, stellar
dynamics and quasi-linear aspects of the growth of large scale
structure in the universe.  In many systems of interest, however,
non-linear interactions play a vital role.  This ultimately requires
the use of self-gravitating $N$-body simulations.  

Fundamentally, solving gravity means solving Poisson's equation for
the gravitational potential, $\phi$, given a mass density, $\rho$:
$\nabla^2 \phi = 4 \pi G \rho$ where $G$ is the Newtonian
gravitational constant.  In a simulations with discrete bodies it is
common to start from the explicit expression for the acceleration,
$a_i = \nabla \phi$ on a given body in terms of the sum of the
influence of all other bodies, $a_i = \sum_{i\neq j} G
M_j/(r_i-r_j)^2$ where the $r_i$ and $M_i$ are the position and masses
of the bodies respectively.  When
attempting to model collisionless systems, these same equations are
the characteristics of the collisionless Boltzmann equation, and the
bodies can be thought of as samples of the distribution function.  In
practical work it is essential to soften the gravitational force on
some scale $r < \epsilon$ to avoid problems with the integration and
to minimize two-body scattering in cases where the bodies
represent a collisionless system.

Early $N$-body work such as studies of relatively small stellar
systems were approached using a direct summation of the forces on each
body due to every other body in the system \cite{aarseth}.  This direct
$\mathcal{O}(N^2)$ approach is impractical for large numbers of bodies, $N$, but has
enjoyed a revival due to incredible throughput of special purpose
hardware such as GRAPE \cite{grape}.  The GRAPE hardware performs
the mutual force calculation for sets of bodies entirely in
hardware and remains competitive with other methods on
more standard floating hardware up to $N\sim100,000$.

A popular scheme for larger $N$ is the Particle-Mesh (PM) method which
has long been used in electrostatics and plasma physics.  The adoption
of PM was strongly related to the realization of the existence of the
$\mathcal{O}(N\,log\,N)$ Fast Fourier Transform (FFT) in the 1960's.
The FFT is used to solve for the gravitational potential from the
density distribution interpolated onto a regular mesh.  In
astrophysics sub-mesh resolution is often desired, in which case the
force can be corrected on sub-mesh scales with local direct sums as in
the Particle-Particle Particle-Mesh (P$^3$M) method.  PM is popular in
stellar disk dynamics, and P$^3$M has seen widespread adoption in
cosmology \citeaffixed{efstathiou}{e.g.}.  PM codes have similarities with
multigrid \citeaffixed{press}{e.g.} and other iterative schemes.
However, working in Fourier space is not only more efficient, but it
also allows efficient force error control through optimization of the
Green's function and smoothing.  Fourier methods are widely recognised
as ideal for large, fairly homogeneous, periodic gravitating
simulations.  Multigrid has some advantages in parallel due to the
local nature of the iterations.  The Particle-Particle correction can
get expensive when particles cluster in a few cells.  Both multigrid
\citeaffixed{flash,kravtsov}{e.g.} and P$^3$M \citeaffixed{ap3m}{AP$^3$M:} can
adapt to address this via a hierarchy of sub-meshes.  With this approach
the serial slow down due to heavy clustering tends toward a fixed
multiple of the unclustered run speed.

In applications such as galactic dynamics where high resolution in
phase space is desirable and particle noise is problematic, the
smoothed gravitational potentials provided by an expansion in modes is
useful.  PM does this with Fourier modes; however, a more elegant
approach is the Self-Consistent Field method (SCF)
\cite{HO,weinberg}.  Using a basis set closely matched to the
evolving system dramatically reduces the number of modes to be
modelled; however, the system must remain close to axi-symmetric and
similar to the basis.  SCF parallelizes well and is also used to
generate initial conditions such as stable individual galaxies
that might be used for merger simulations.

A current popular choice is to use tree algorithms which are
inherently $\mathcal{O}(N\,log\,N)$.  This approach recognises that
details of the remote mass distribution become less important for
accurate gravity with increasing distance.  Thus the remote mass
distribution can be expanded in multipoles on the different size
scales set by a tree-node hierarchy.  The appropriate scale to use is
set by the opening angle subtended by the tree-node bounds relative to
the point where the force is being calculated.  The original
Barnes-Hut \cite{BH86} method employed oct-trees but this is not
especially advantageous, and other trees also work well \cite{porter}.
The tree approach can adapt to any topology, and thus the speed of the
method is somewhat insensitive to the degree of clustering.  Once a
tree is built it can also be re-used as an efficient search method for
other physics such as particle based hydrodynamics.

A particularly useful property of tree codes is the ability to
efficiently calculate forces for a subset of the bodies.  This is
critical if there is a large range of time-scales in a simulation and
multiple independent timesteps are employed.  At the cost of force
calculations no longer being synchronized among the particles
substantial gains in time-to-solution may be realized.
Multiple timesteps are particularly important for current
astrophysical applications where the interest and thus resolution
tends to be focused on small regions within large simulated environments
such as individual galaxies, stars or planets.  Dynamical times can
become very short for small numbers of particles.  P$^3$M codes are
faster for full force calculations but are difficult to adapt to
calculate a subset of the forces.

In order to treat periodic boundaries with tree codes it is necessary
to effectively infinitely replicate the simulation volume which may be
approximated with an Ewald summation \cite{hbs91}.  An efficient
alternative which is seeing increasing use is to use trees in place of
the direct Particle-Particle correction to a Particle-Mesh code, often
called Tree-PM \cite{wadsleytreepm,bode,bagla}.

The Fast Multipole Method (FMM) recognises that the applied force as
well as the mass distribution may be expanded in multipoles.  This
leads to a force calculation step that is $\mathcal{O}(N)$ as each
tree node interacts with a similar number of nodes independent of $N$
and the number of nodes is proportional to the number of bodies.
Building the tree is still $\mathcal{O}(N\,log\,N)$ but this is a
small cost for simulations up to $N\sim 10^7$ \cite{dehnen}.  The
\citeasnoun{GR} method used spherical harmonic expansions where the desired
accuracy is achieved solely by changing the order of the expansions.
For the majority of astrophysical applications the allowable force
accuracies make it much more efficient to use fixed order Cartesian
expansions and an opening angle criterion similar to standard tree
codes \cite{SW,dehnen}.  This approach has the nice property of
explicitly conserving momentum (as do PM and P$^3$M codes).  The
prefactor for Cartesian FMM is quite small so that it can outperform
tree codes even for small $N$ \cite{dehnen}.  It is a significantly
more complex algorithm to implement, particularly in parallel.  One
reason widespread adoption has not occurred is that the speed benefit
over a tree code is significantly reduced when small subsets of the
particles are having forces calculated (e.g. for multiple timesteps).

\section{Solving Gravity in Gasoline}\label{grav}

Gasoline is built on the Pkdgrav framework and thus uses the same
gravity algorithms.  The Pkdgrav parallel $N$-body code was designed
by Stadel and developed in conjunction with Quinn beginning in the
early 90's.  This includes the parallel code structure and core
algorithms such as the tree structure, tree walk, hexadecapole
multipole calculations for the forces and the Ewald summation.  There
have been additional contributions to the gravity code by Richardson
and Wadsley in the form of minor algorithmic modifications and
optimizations.  The latter two authors became involved as part of the
collisions and Gasoline extensions of the original Pkdgrav code
respectively.  We have summarized the gravity method used in the
production version of Gasoline without getting into great mathematical
detail.  For full technical details on Pkdgrav the reader is referred
to \citeasnoun{stadel}.

\subsection{Design}\label{design}

Gasoline is fundamentally a tree code.  It uses a variant on the K-D
tree (see below) for the purpose of calculating gravity, dividing work
in parallel and searching.  \citeasnoun{stadel} designed Pkdgrav from the
start as a parallel code.  There are four layers in the code.  The Master
layer is essentially serial code that orchestrates overall progress of
the simulation.  The Processor Set Tree (PST) layer distributes work
and collects feedback on parallel operations in an architecture
independent way using MDL. The Machine Dependent Layer (MDL) is a
relatively short section of code that implements remote procedure
calls, effective memory sharing and parallel diagnostics.  All processors
other than the master loop in the PST level waiting for directions from the
single process executing the Master level.  Directions are passed down
the PST in a tree based $\mathcal{O}(log_2 N_P)$ procedure that ends
with access to the fundamental bulk particle data on every node at the
PKD level.  The Parallel K-D (PKD) layer is almost entirely serial but
for a few calls to MDL to access remote data.  The PKD layer manages
local trees for gravity and particle data and is where the physics is
implemented.  This modular design enables new physics to be coded at
the PKD level without requiring detailed knowledge of the parallel
framework.

\subsection{Mass Moments}

Pkdgrav departed significantly from the original $N$-body tree code
designs of \citeasnoun{BH86} by using 4th (hexadecapole) rather than 2nd
(quadrupole) order multipole moments to represent the mass
distribution in cells at each level of the tree.  This results in less
computation for the same level of accuracy: better pipelining, smaller
interaction lists for each particle and reduced communication demands
in parallel.  The current implementation in Gasoline uses reduced
moments that require only $n+1$ terms to be stored for the $n^{th}$
moment.  For a detailed discussion of the accuracy and efficiency of
the tree algorithm as a function the order of the multipoles used see
\cite{stadel} and \cite{SW}.

\subsection{The Tree}

The original K-D tree \cite{bentley} was a balanced binary tree.
Gasoline divides the simulation in a similar way using recursive
partitioning.  At the PST level this is parallel domain decomposition
and the division occurs on the longest axis to recursively divide the
work among the remaining processors.  Even divisions occur only when
an even number of processors remains.  Otherwise the work is split in
proportion to the number of processors on each side of the division.
Thus,  Gasoline may use arbitrary numbers of processors and is efficient for
flat topologies without adjustment.  At the PST level
each processor has a local rectangular domain within which a local
binary tree is built.  The structure of the lower tree is 
important for the accuracy of gravity and efficiency of other search
operations such as neighbour finding required for SPH.

Oct-trees \citeaffixed{BH86,SW}{e.g.} are traditional in gravity codes.
In contrast, the key data structures used by Gasoline are spatial
binary trees.  One immediate gain is that the local trees do not have
to respect a global structure and simply continue from the PST level
domain decomposition in parallel.  The binary tree determines the
hierarchical representation of the mass distribution with multipole
expansions, of which the root node or cell encloses the entire
simulation volume.  The local gravity tree is built by recursively
bisecting the longest axis of each cell which keeps the cells axis
ratios close to one.  In contrast, cells in standard K-D trees can
have large axis ratios which lead to large multipoles and with
correspondingly large gravity errors.  At each level the dimensions of
the cells are squeezed to just contain the particles.  This overcomes
the empty cell problem of un-squeezed spatial bisection trees.  The
SPH tree is currently a balanced K-D tree; however, testing indicates
that the efficiency gain is slight and it is not worth the cost of an
additional tree build.

The top down tree build process is halted when $n_{Bucket}$ or fewer
particles remain in a cell.  Stopping the tree with $n_{Bucket}$
particles in a leaf cell reduces the storage required for the tree and
makes both gravity and search operations more efficient.  For these
purposes $n_{Bucket} \sim 8-16$ is a good choice.

Once the gravity tree has been built there is a bottom-up pass
starting from the buckets and proceeding to the root, calculating the
center of mass and the multipole moments of each cell from the center
of mass and moments of each of its two sub-cells.

\subsection{The Gravity Walk} 

Gasoline calculates the gravitational accelerations using the well
known tree-walking procedure of the \citeasnoun{BH86} algorithm,
except that it collects interactions for entire buckets rather than
single particles.  This amortizes the cost of tree traversal for a
bucket over all its particles.

In the tree building phase, Gasoline  assigns to each  cell of the tree
an {\em opening radius} about its center-of-mass. This is defined as,
\begin{equation}
 r_{\rm open} = {2 B_{\rm max} \over \sqrt{3} \; \theta}
\label{eq:1}
\end{equation}
where $B_{\rm max}$ is the maximum  distance
from a  particle in the  cell to the center-of-mass of the cell.
The {\em opening angle}, $\theta$, is a user specified accuracy parameter which is
similar to  the traditional  $\theta$  parameter of  the Barnes-Hut code;
notice  that   decreasing $\theta$ in equation   \ref{eq:1}, increases
$r_{\rm open}$.

\begin{figure*}
\epsfbox{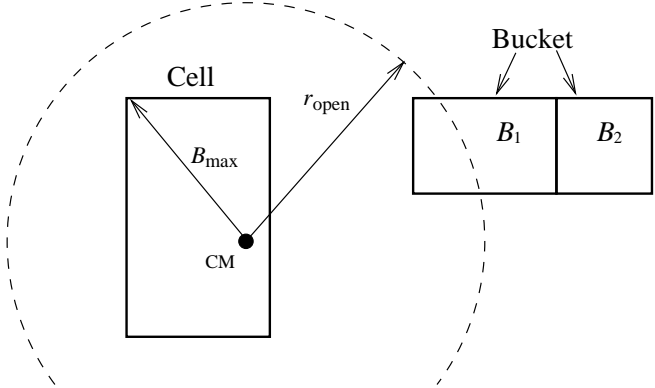}
\caption{Opening radius for a cell in the tree, intersecting
bucket $B_1$ and not bucket $B_2$. This cell is ``opened'' when
walking the tree for $B_1$. When walking the tree for $B_2$, the cell
will be added to the particle-cell interaction list of $B_2$.}
\label{fig:2.2}
\end{figure*}

The opening radii are used in the {\em Walk\/} phase of the algorithm as
follows: for each bucket  $B_i$, Gasoline starts  descending the
tree, opening those  cells whose $r_{\rm open}$ intersect with
$B_i$   (see Figure \ref{fig:2.2}).  If   a cell is opened,
then Gasoline repeats  the intersection-test with  $B_i$ for the
cell's   children    Otherwise, the  cell   is   added  to  the  {\em
particle-cell interaction list} of $B_i$.  When Gasoline reaches
the leaves of the tree and a {\em bucket}  $B_j$ is opened, all
of $B_j$'s particles are added  to  the {\em particle-particle
interaction} list of $B_i$.

Once the tree has been  traversed in this  manner we can calculate the
gravitational acceleration  for each  particle   of $B_i$  by
evaluating the interactions specified  in the two lists.  Gasoline uses
the hexadecapole multipole expansion to calculate particle-cell interactions.

\subsection{Softening}

The particle-particle interactions are softened to lessen two-body
relaxation effects that compromise the attempt to model continuous
fluids, including the collisionless dark matter fluid.  In Gasoline
the particle masses are effectively smoothed in space using the same
spline form employed for SPH in section~\ref{SPH}.  This means that
the gravitational forces vanish at zero separation and return to
Newtonian $1/r^2$ at a separation of $\epsilon_i+\epsilon_j$ where
$\epsilon_i$ is the gravitational softening applied to each particle.
In this sense the gravitational forces are well matched to the SPH
forces.

\subsection{Periodicity}

A disadvantage of tree codes is that they must deal explicitly with
periodic boundary conditions (as are usually required for cosmology).
Gasoline incorporates periodic boundaries via the Ewald summation
technique where the force is divided into short and long range
components.  The Gasoline implementation differs from that of
\citeasnoun{hbs91} in using a new technique due to \citeasnoun{stadel} based on
an hexadecapole moment expansion of the fundamental cube to
drastically reduce the work for the long range Ewald sum that
approximates the infinite periodic replications.  For each particle
the computations are local and fixed, and thus the algorithm scales
exceedingly well.  There is still substantial additional work in
periodic simulations because particles interact with cells and
particles in nearby replicas of the fundamental cube \citeaffixed{ding}{similar to}.

\subsection{Force Accuracy}\label{accuracy}

\begin{figure}[h]
\epsfxsize=5.5in \epsfbox{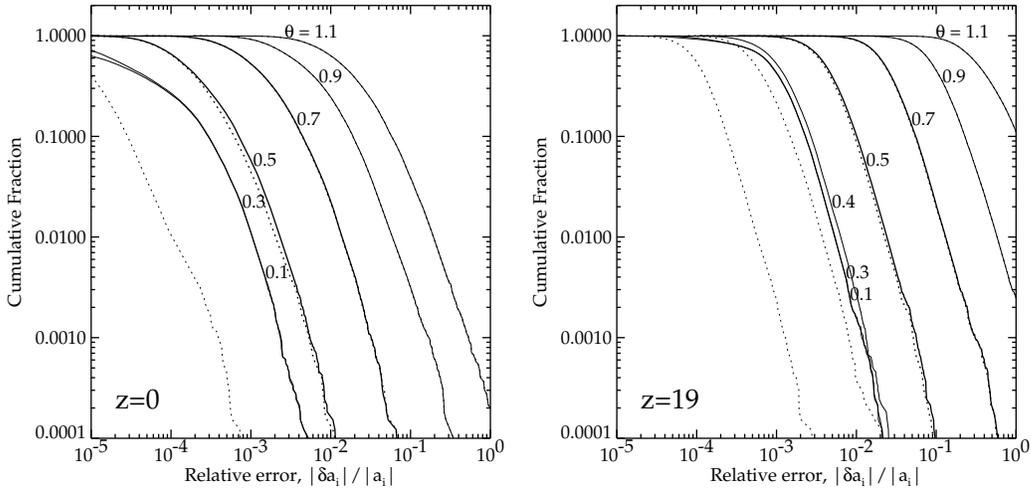}
\caption{Gasoline relative errors, for various opening angles
$\theta$.  The distributions are all for a $32^3$, 64 Mpc box where
the left panel represents the clustered final state and the right an
initial condition (redshift $z=19$).  Typical
values of $\theta=0.5$ and $0.7$ are shown as thick lines.  The solid
curves compare to exact forces and thus have a minimum error set by
the parameters of the Ewald summation whereas the dotted curves
compare to the $\theta=0.1$ case.  The Ewald summation errors only
become noticeable for $\theta < 0.4$.  Relative errors are best
for clustered objects but over-emphasize errors for low acceleration
particles that are common in cosmological initial conditions.}
\label{accfig}
\end{figure}

\begin{figure}
\epsfxsize=5.5in
\epsfbox{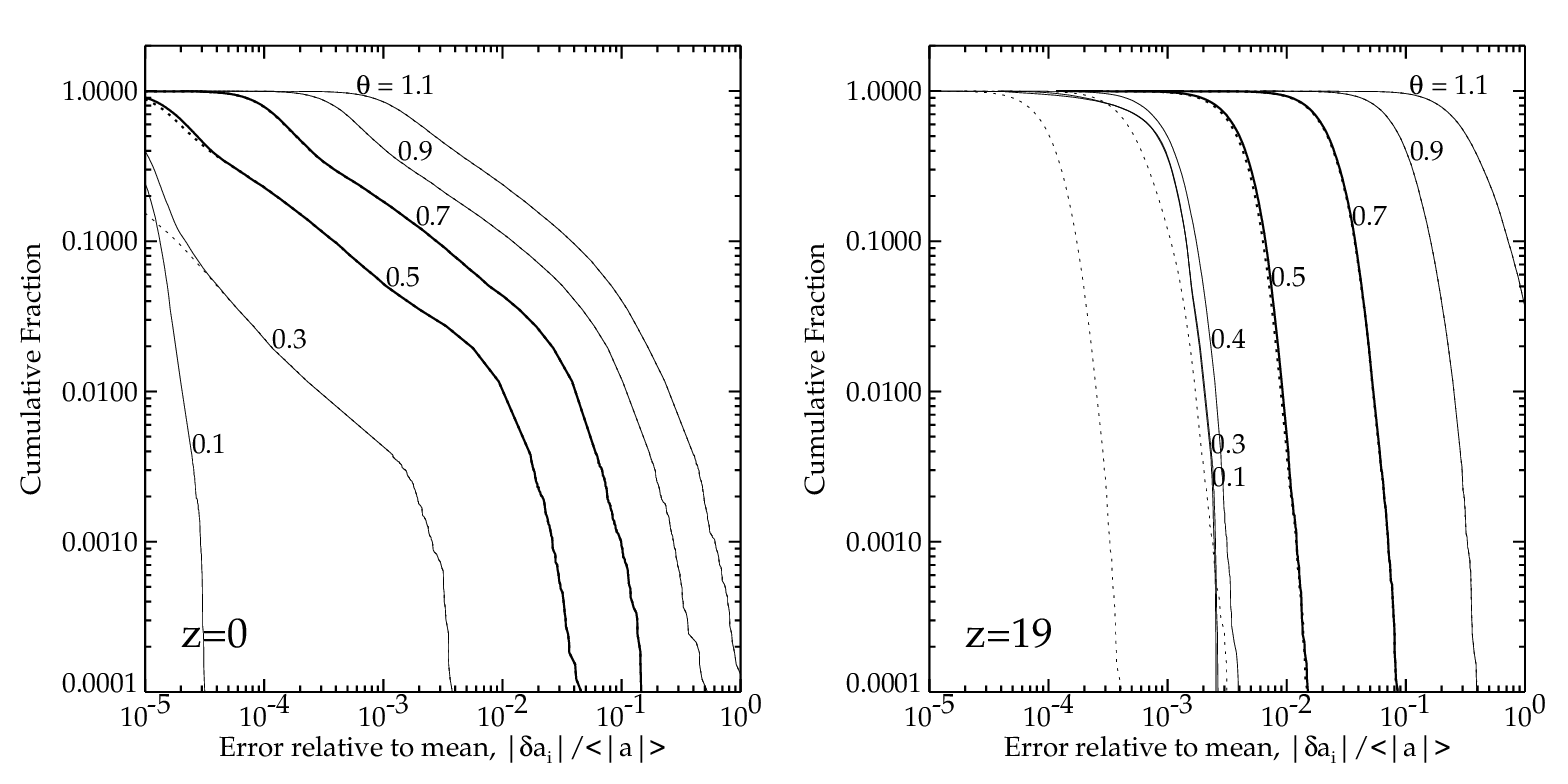}
\caption{Gasoline errors relative to the mean for the same cases as
Fig.~\ref{accfig}.  The errors are measured compared to the mean
acceleration (magnitude) averaged over all the particles.  Errors
relative to the mean are more appropriate to judge accuracy for small
accelerations resulting from force cancellation in a nearly
homogeneous medium.}
\label{accfigabs}
\end{figure}

The tree opening criteria places a bound on the relative error due to
a single particle-cell interaction.  As Gasoline uses hexadecapoles the
error bound improves rapidly as the opening angle, $\theta$, is
lowered.  The relationship between the typical (e.g. rms) relative
force error and opening angle is not a straight-forward power-law in
$\theta$ because the net gravitational forces on each particle result
from the cancellation of many opposing forces.  In
figure~\ref{accfig}, we show a histogram of the relative acceleration
errors for a cosmological Gasoline simulations at two different epochs
for a range of opening angles.  We have plotted the error curves as cumulative
fractions to emphasize the limited tails to high error values.
For typical gasoline simulations we
commonly use $\theta=0.7$ which gives an {\it rms} relative error of
0.0039 for the clustered final state referred to by the left panel of
figure~\ref{accfig}.  The errors relative to the mean acceleration
(figure~\ref{accfigabs}) are larger (rms 0.0083) but of less interest
for highly clustered cases.

As a simulated medium becomes more uniform the net gravitational
accelerations approach zero.  In a tree code the small forces result
from the cancellation of large opposing forces.  This is
the case in cosmology at early times when the perturbations ultimately
leading to cosmological structures are still small.  In this situation it
is essential to tighten the cell-opening criterion to increase the
relative accuracy so that the net accelerations are sufficiently
accurate.  For example in the right hand panels of
figures~\ref{accfig} and~\ref{accfigabs} the errors are larger, and
at $z=19$, the {\it rms} relative error is 0.041 for
$\theta=0.7$.  However, here the absolute errors are lower by nearly a
factor of two in rms (0.026) as shown in figure~\ref{accfigabs}.  At
early times when the medium fairly homogeneous growth is driven by
local gradients, and thus acceleration errors normalized to the
mean acceleration provide the better measure of accuracy as large relative
errors are meaningless for accelerations close to zero.  To ensure
accurate integrations we switch to values such as $\theta=0.5$ before
$z=2$, giving an {\it rms} relative error of 0.0077 and an {\it rms}
error of 0.0045 normalized to the mean absolute acceleration for the
$z=19$ distribution (a reasonable start time for a cosmological
simulation on this scale).

\section{Gasdynamics}\label{gas}

Astrophysical systems are predominantly at very low physical
densities and experience wide-ranging temperature variations.  Most of
the material is in a highly compressible gaseous phase.  In general
this means that a perfect adiabatic gas is an excellent approximation
for the system.  Microscopic physical processes such as shear viscosity and
diffusion can usually be neglected.  High-energy processes and the
action of gravity tend to create large velocities so that flows are
both turbulent and supersonic: strong shocks and very high Mach
numbers are common.  Radiative cooling processes can also be
important; however, the timescales can often be much longer or shorter
than dynamical timescales.  In the latter case isothermal gas is often
assumed for simplicity.  In many areas of computational astrophysics,
particularly cosmology, gravity tends to be the dominating force that
drives the evolution of the system.  Visible matter, usually in the
form of radiating gas, provides the essential link to observations.
Radiative transfer is always present but may not significantly affect
the energy and thus pressure of the gas during the simulation.

Fluid dynamics solvers can be broadly classified into Eulerian or
Lagrangian methods.  Eulerian methods use a fixed computational mesh
through which the fluid flows via explicit advection terms.  Regular
meshes provide for ease of analysis and thus high order methods such
as PPM \cite{woodcoll} and TVD schemes \citeaffixed{harten,kurganov}{e.g.} have been
developed.  The inner loops of mesh methods can often be pipelined for
high performance.  Lagrangian methods follow the evolution of fluid
parcels via the full (comoving) derivatives.  This requires a
deforming mesh or a mesh-less method such as Smoothed Particle
Hydrodynamics (SPH) \cite{M92}.  Data management is more complex in
these methods; however, advection is handled implicitly and the
simulation tends to naturally adapt to follow density contrasts.

Large variations in physical length scales in astrophysics have
limited the usefulness of Eulerian grid codes.  Adaptive Mesh
Refinement (AMR) \cite{flash,bryan} overcomes this at the
cost of data management overheads and increased code complexity.  In
the cosmological context there is the added complication of dark
matter.  There is more dark matter than gas in the universe so it
dominates the gravitational potential.  Perturbations present on all
scales in the dark matter guide the formation of gaseous structures
including galaxies and the first stars.  A fundamental limit to AMR in
computational cosmology is matching the AMR resolution to the
underlying dark matter resolution.  Particle based Lagrangian methods
such as SPH are well matched to this constraint.  A useful feature of
Lagrangian simulations is that bulk flows (which can be highly
supersonic in the simulation frame) do not limit the timesteps.
Particle methods are well suited to rapidly rotating systems such as
astrophysical disks where arbitrarily many rotation periods may have
to be simulated (e.g. SPH explicitly conserves angular momentum).  A
key concern for all methods is correct angular momentum transport.

\section{Smoothed Particle Hydrodynamics in Gasoline}\label{SPH}

Smoothed Particle Hydrodynamics is an approach to hydrodynamical
modelling developed by \citeasnoun{lucy} and \citeasnoun{gm1977}. It is a
particle method that does not refer to grids for the calculation of
hydrodynamical quantities: all forces and fluid properties are found
on moving particles eliminating numerically diffusive advective terms.
The use of SPH for cosmological simulations required the development
of variable smoothing to handle huge dynamic ranges \cite{hk}.  SPH
is a natural partner for particle based gravity.  SPH has been
combined with P$^3$M \cite{evrard}, Adaptive P$^3$M \citeaffixed{hydra}{HYDRA,},
GRAPE \cite{steinmetz} and tree gravity \cite{hk}.  Parallel codes
using SPH include Hydra MPI,  Parallel TreeSPH
\cite{dave} and the GADGET tree code \cite{springel}.

The basis of the SPH method is the representation and evolution of
smoothly varying fluid quantities whose value is only known at
disordered discrete points in space occupied by particles.  Particles
are the fundamental resolution elements comparable to cells in a mesh.
SPH functions through local summation operations over particles
weighted with a smoothing kernel, $W$, that approximates a local
integral.  The smoothing operation provides a basis from which to
obtain derivatives.  Thus, estimates of density related physical
quantities and gradients are generated.  The summation aspect led to
SPH being described as a Monte Carlo type method (with
$\mathcal{O}(1/\sqrt{N})$ errors) however it was shown by
\citeasnoun{m1985a} that the method is more closely related to
interpolation theory with errors $\mathcal{O}((ln N)^d/N)$, where $d$
is the number of dimensions.

A general smoothed estimate for some quantity $f$ at particle $i$
given particles $j$ at positions $\vec{r}_j$ takes the form:
\begin{eqnarray}
f_{i,{\rm smoothed}} = \sum_{j=1}^{n} f_j W_{ij}(\vec{r}_i-\vec{r}_j,h_i,h_j),
\label{sum}
\end{eqnarray}
\noindent where $W_{ij}$ is a kernel function and $h_j$ is a smoothing length
indicative of the range of interaction of particle $j$.  It is common
to convert this particle-weighted sum to volume weighting using $f_j
m_j/\rho_j$ in place of $f_j$ where the $m_j$ and $\rho_j$ are the
particle masses and densities, respectively.  For momentum and energy
conservation in the force terms, a symmetric $W_{ij}=W_{ji}$ is
required.  We use the kernel-average first suggested by \citeasnoun{hk},

\begin{eqnarray}
W_{ij}=\frac{1}{2}
w(|\vec{r}_i-\vec{r}_j|/h_i) + \frac{1}{2} w(|\vec{r}_i-\vec{r}_j|/h_j) 
\label{kernelavg}
\end{eqnarray}

For $w(x)$ we use the standard spline form with compact support
where $w=0$ if $x>2$ \cite{M92}.

We employ a fairly standard implementation of the the hydrodynamics
equations of motion for SPH \cite{M92}.  Density is calculated from a
sum over particle masses $m_j$,
\begin{eqnarray}
\rho_i=\sum_{j=1}^{n} m_j W_{ij}.
\label{denssum}
\end{eqnarray}
The momentum equation is expressed,
\begin{eqnarray}
\frac{d\vec{v}_i}{dt}& = & -\sum_{j=1}^{n}m_j 
\left({\frac{P_i}{\rho_i^2}+\frac{P_j}{\rho_j^2}+\Pi_{ij}}\right)
\nabla_i W_{ij}, 
\label{sphmom}
\end{eqnarray}
\noindent where $P_j$ is pressure, $\vec{v}_i$ velocity and the artificial viscosity term $\Pi_{ij}$ is given by,
\begin{eqnarray}
\Pi_{ij} = \left\{{ \begin{array}{ll}
\frac{-\alpha\frac{1}{2}(c_i+c_j)\mu_{ij}+\beta\mu_{ij}^2}{\frac{1}{2} (\rho_{i}+\rho{j})}
& {\rm for~}\vec{v}_{ij}\cdot\vec{r}_{ij}< 0,\\
~0 & {\rm otherwise}, \end{array}}\right.\\
{\rm\ where\ }
\mu_{ij} = \frac{h(\vec{v}_{ij}\cdot\vec{r}_{ij})}{\vec{r}_{ij}^{\,2}+0.01 (h_i+h_j)^2},
\label{artifvisc}
\end{eqnarray}
\noindent where $\vec{r}_{ij}=\vec{r}_i-\vec{r}_j$,
$\vec{v}_{ij}=\vec{v}_i-\vec{v}_j$ and $c_j$ is the sound speed.
$\alpha=1$ and $\beta=2$ are coefficients we use for the terms
representing shear and Von Neumann-Richtmyer (high Mach number)
viscosities respectively.  When simulating strongly rotating systems
we use the multiplicative \citeasnoun{balsara} switch $\frac{\mid\nabla\cdot\vec{v}\mid}
{\mid\nabla\cdot\vec{v}\mid+\mid\nabla\times\vec{v}\mid}$ 
to suppress the viscosity in non-shocking, shearing
environments.

The pressure averaged energy equation (analogous to equation~\ref{sphmom})
conserves energy exactly in the limit of infinitesimal time steps but
may produce negative energies due to the $P_j$ term if significant
local variations in pressure occur.  We employ the following equation
\citeaffixed{evrard,benz}{advocated by} which also conserves energy exactly in each
pairwise exchange but is dependent only on the local particle
pressure,

\begin{eqnarray}
\frac{d\ u_i}{dt}& = & \frac{P_i}{\rho_i^2} \sum_{j=1}^{n}{m_j} 
\vec{v}_{ij} \cdot \nabla_i W_{ij}, 
\end{eqnarray}

where $u_i$ is the internal energy of particle $i$, which is equal to
$1/(\gamma-1) P_i/\rho_i$ for an ideal gas.  In this formulation entropy is
closely conserved making it similar to alternative entropy integration approaches
such as that proposed by \citeasnoun{sh}.

SPH derivative estimates, such as the rate of change of thermal
energy, vary sufficiently from the exact answer so that
over a full cosmological simulation the cooling due to universal
expansion will be noticeably incorrect.  In this case we use comoving
divergence estimates and add the cosmological expansion term
explicitly.

\subsection{Neighbour Finding}\label{nbr}

Finding neighbours of particles is a useful operation.  A neighbour
list is essential to SPH, but it can also be a basis for other local estimates,
such as a dark matter density, as a first step in finding potential
colliders or interactors via a short range force.  Stadel developed an
efficient search algorithm using priority queues and a K-D tree
ball-search to locate the $k$-nearest neighbours for each particle
(freely available as the Smooth utility at
\url{http://www-hpcc.astro.washington.edu}).  For Gasoline we use a
parallel version of the algorithm that caches non-local neighbours via
MDL.  The SPH interaction distance $2 h_i$ is set equal to the $k$-th
neighbour distance from particle $i$.  We use an exact number of
neighbours.  The number is set to a value such as 32 or 64 at start-up.  We have also
implemented a minimum smoothing length which is usually set to be
comparable to the gravitational softening.
 
To calculate the fluid accelerations using SPH we perform two smoothing
operations.  First we sum densities and then forces using the density
estimates.  To get a kernel-averaged sum for every particle
(equations \ref{sum},\ref{kernelavg}) it is sufficient to perform a gather
operation over all particles within $2 h_i$ of every particle $i$.  If
only a subset of the particles are active and require new forces, all
particles for which the active set are neighbours must also perform a
gather so that they can scatter their contribution to the active set.
Finding these scatter neighbours requires solving the $k$-inverse
nearest neighbour problem, an active research area
in computer science \citeaffixed{anderson}{e.g.}.  Fortunately, during a simulation the change
per step in $h$ for each particle typically less than 2-3 percent, so
it is sufficient to find scatter neighbours, $j$ for which some active
particles is within $2 h_{j,OLD}(1+e)$.  We use a tree search where
the nodes contain SPH interaction bounds for their particles estimated
with $e=0.1$.  A similar scheme has been employed by \citeasnoun{springel}.
For the forces sum the inactive neighbours
need density values which can be estimated using the continuity
equation or calculated exactly with a second inverse neighbour
search.

\subsection{Cooling}\label{cooling}

In astrophysical systems the cooling timescale is usually short
compared to dynamical timescales which often results in temperatures
that are close to an equilibrium set by competing heating and cooling
processes.  We have implemented a range of cases including: adiabatic
(no cooling), isothermal (instant cooling), and implicit energy
integration.  Hydrogen and Helium cooling processes have been
incorporated.  Ionization fractions are calculated assuming
equilibrium for efficiency.  Gasoline optionally adds heating due to
feedback from star formation, an uniform UV background
or using user defined functions.

The implicit integration uses a stiff equation solver assuming that
the hydrodynamic work and density are constant across the step.  The
second order nature of the particle integration is maintained using an
implicit predictor step when needed.  Internal energy is required on
every step to calculate pressure forces on the active particles.  The
energy integration is $\mathcal{O}(N)$ but reasonably expensive.  To
avoid integrating energy for every particle on the smallest timestep
we extrapolate each particle forward on its individual dynamical
timestep and use linear interpolation to estimate internal energy at
intermediate times as required.  

\section{Integration: Multiple Timesteps}\label{stepping}

The range of physical scales in astrophysical systems is large.  For
example current galaxy formation simulations contain 9 orders of
magnitude variation in density.  The dynamical timescale for gravity
scales as $\rho^{-1/2}$ and for gas it scales as $\rho^{-1/3}\
T^{-1/2}$.  For an adiabatic gas the local dynamical time scales
as $\rho^{-2/3}$.  With gas cooling (or the isothermal assumption)
simulations can achieve very high gas densities.  In most cases gas
sets the shortest dynamical timescales, and thus gas simulations are
much more demanding (many more steps to completion) than corresponding
gravity only simulations.  Time adaptivity can be very effective in
this regime.

\begin{figure}
\epsfxsize=5.5in
\epsfbox{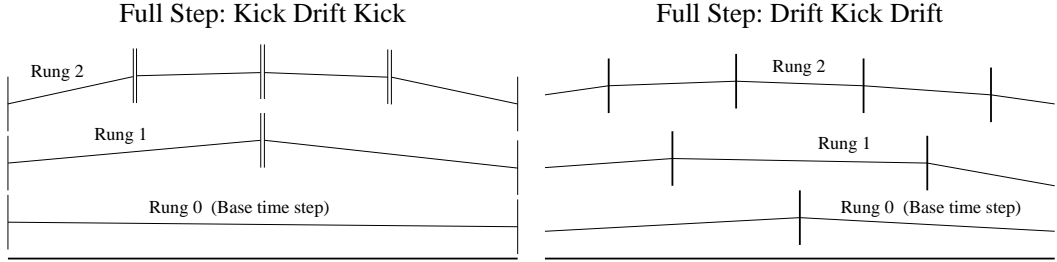}
\caption{Illustration of multiple timestepping: the linear sections represent particle positions during a Drift step with time along the x-axis.  The vertical bars represent Kicks changing the velocity.  In KDK, the accelerations are applied as two half-kicks (from only one force evaluation) which is why there are two bars
interior of the KDK case. At the end of each full step all variables are synchronised.}
\label{timestep}
\end{figure}

Gasoline incorporates the timestep scheme described as
Kick-Drift-Kick (KDK) in \citeasnoun{Quinn97}.  The scheme uses a fixed
overall timestep.  Starting with all quantities synchronised,
velocities and energies are updated to the half-step (half-Kick),
followed by a full step position update (Drift).  The positions alone
are sufficient to calculate gravity at the end of the step; however, for
SPH, velocities and thermal energies are also required and obtained
with a predictor step using the old accelerations.  Finally another
half-Kick is applied synchronising all the variables.  Without gas forces
this is a symplectic leap-frog integration.  The leap-frog scheme
requires only one force evaluation and minimum storage.  While symplectic
integration is not required in cosmology it is critical in
solar system integrations or any system where natural instabilities
occur over very many dynamical times.

An arbitrary number of sub-stepping rungs factors of two smaller may
be used as shown in Figure~\ref{timestep}.  The scheme is no longer
strictly symplectic if particles change rungs during the integration
which they generally must do to satisfy their individual timestep
criteria.  After overheads, tree-based force calculation scales
approximately with the number of active particles so large speed-ups
may be realised in comparison to single stepping (see
section~\ref{performance}).  Figure~\ref{timestep} compares KDK with
Drift-Kick-Drift (DKD), also discussed in \citeasnoun{Quinn97}.  For KDK
the force calculations for different rungs are synchronised.  In the
limit of many rungs this results in half as many force calculation
times with their associated tree builds compared to DKD.  KDK also
gives significantly better momentum and energy conservation for the
same choice of timestep criteria.

We use standard timestep criteria based on the particle acceleration,
and for gas particles, the Courant condition and the expansion cooling rate.

\begin{eqnarray}
dt_{Accel} \ & \leq \ & 0.3 \ \sqrt{\frac{a}{\epsilon}} \nonumber \\
dt_{Courant} \ & \leq \ & 0.4 \ \frac{h}{(1+\alpha)\,c + \beta\,\mu_{MAX}} \nonumber \\
dt_{Expand} \ & \leq \ & 0.25 \ \frac{u}{du/dt} \ \  {\rm if }\  du/dt < 0 
\end{eqnarray}

$\mu_{MAX}$ is the maximum value of $|\mu_{ij}|$ (from
equation~\ref{artifvisc}) over interactions between pairs of SPH particles.

For cosmology in place of the comoving velocity $\vec{v}$, we use the
momentum $\vec{p} = a^2 \vec{v}$ which is canonical to the comoving
position, $\vec{x}$.  As described in detail in Appendix A of
\cite{Quinn97}, this results in a separable Hamiltonian which may be
integrated straightforwardly using the Drift and Kick operators,
\begin{eqnarray}
{\rm Drift}\  D(\tau): \vec{x}_{t+\tau} = \vec{x}_t + \vec{p} \int_t^{t+\tau} \frac{dt}{a^2}, \ \
{\rm Kick}\  K(\tau): \vec{p}_{t+\tau} = \vec{p}_t + \nabla \phi \int_t^{t+\tau} \frac{dt}{a},
\end{eqnarray}
where $\phi$ is the perturbed potential given by $\nabla^2 \phi = 4
\pi G a^2 (\rho-\bar{\rho})$, $a$ is the cosmological expansion factor
and $\bar{\rho}$ is the mean density.  Thus no Hubble drag term is
required in the equations of motion, and the integration is perfectly
symplectic in the single stepping case.

\section{Performance}\label{performance}

\begin{figure}[h]
\epsfxsize=5.5in
\epsfbox{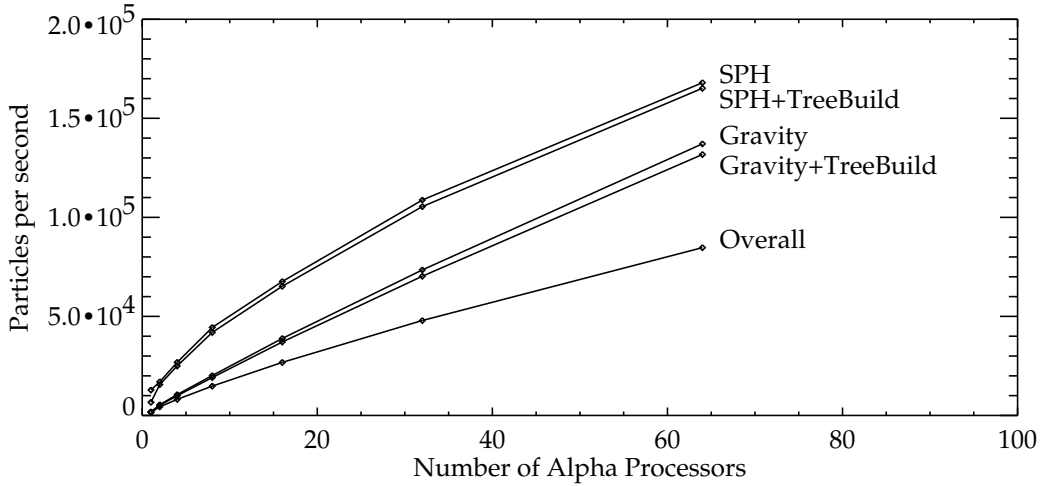}
\caption{Science rate in particles per second for Gasoline in
parallel.  Scaling tests were performed on an HP Alphaserver SC (ev67 667
MHz processors).  The timing was done on a single force calculation
and update for a standard cosmological volume in the clustered end
state ($z=0$) with $128^3$ dark and $128^3$ gas particles.  See the
text for more details.}
\label{fig:particlespersecond}
\end{figure}

Gasoline was built on the Pkdgrav $N$-body code which achieves
excellent performance on pure gravity in serial and parallel.
Performance can be measured in floating point calculations per second
but the measure of most interest to researchers is the science rate.
We define this in terms of resolution elements updated per unit
wallclock time.  In the case of Gasoline this is particles updated per
second.  This measure directly determines how long it takes to finish
the simulation.  Figure~\ref{fig:particlespersecond} shows the scaling
of particles per second with numbers of alpha processors for a single
update for velocities and positions for all particles.  This requires
gravitational forces for all particles and SPH for the gas particles (half the
total).  The simulation used is a $128^3$ gas and $128^3$ dark matter
particle, purely adiabatic cosmological simulation ($\Lambda$CDM) in a
200 Mpc box at the final time (redshift, $z=0$).  At this time the
simulation is highly clustered locally but material is still well
distributed throughout the volume.  Thus, it is still possible to
partition the volume to achieve fairly even work balancing among a
fair number of processors.  As a rule of thumb we aim to have around
100,000 particles per processor.  As seen in the figure, the gravity
scales very well.  For pure gravity around $80\%$ scaling efficiency
can be achieved in this case.  The cache based design has small memory
overheads and uses the large amounts of gravity work to
hide the communication overheads associated with filling the cache
with remote particle and cell data.  With gas the efficiency is
lowered because the data structures to be passed are larger and
there is less work per data element to hide the
communication costs.  Thus the computation to communication ratio is
substantially lowered with more processors.  Fixed costs such as the
treebuilds for gravity and SPH scale well, but the gas-only costs 
peel away from the more efficient gravity scaling.  When the overall rate
is considered, Gasoline is still making substantial improvements in the
time to solution going from 32 to 64 processors (lowest curve in the
figure).  The overall rate includes costs for domain decomposition and
$\mathcal{O}(N)$ updates of the particle information such as updating the positions from
the velocities.  The net science
rate compares well with other parallel Tree and SPH codes.
\begin{figure}[h]
\epsfxsize=5.5in
\epsfbox{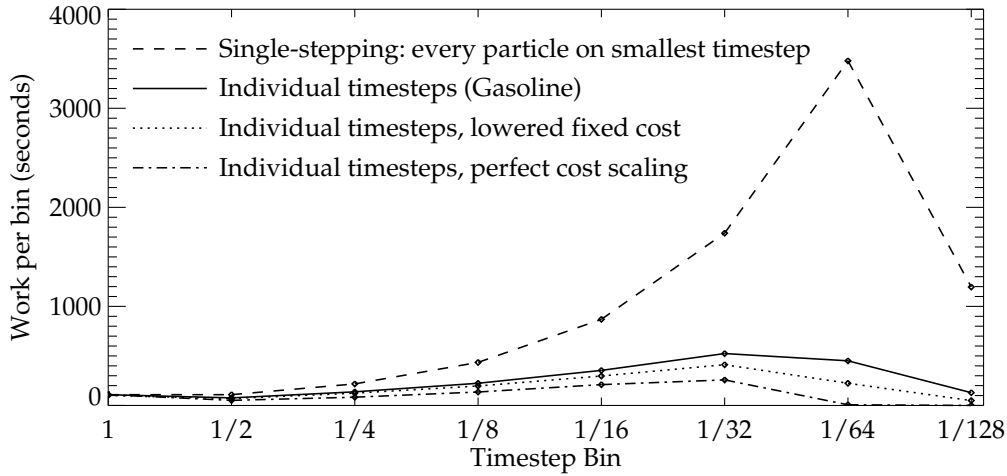}
\caption{Benefits of multistepping.  Each major timestep is performed
as a hierarchical sequence of substeps with varying numbers of
particles active in each timestep bin.  We are using a realistic
case:  a single major step of a 
galaxy simulation at redshift $z=1$ (highly clustered) run on 8 alpha
processors with a range of 128 in timesteps.  Gasoline (solid line)
achieves an overall 4.1 times speed-up over a single stepping code
(dashed line).  Planned improvements in the fixed costs such as tree
building (dotted line) would give a 5.3 times speedup.  If all costs
could be scaled to the number of particles in each timestep bin
(dash-dot line), the result would be a 9.5 times speed-up over single
stepping.}
\label{fig:multistep}
\end{figure}

Figure~\ref{fig:particlespersecond} only treats single timestepping.
Multiple timesteps provide an additional speed-up in the science rate
of a factor typically in the range of 2-5 that is quite problem
dependent.  The value of individual particle timesteps is illustrated
separately in figure~\ref{fig:multistep}.  For this example we
analysed the time spent by Gasoline doing a single major step (around
13 Million years) of a million particle Galaxy formation simulation at
redshift $z=1$.  The simulation used a range of 128 in substeps per
major step.  The top curve in the figure is the cost for single
stepping.  It rises exponentially, doubling with every new bin, and
drops off because the last bin was not always occupied.  Sample
numbers of particles in the bins (going from 0 to 7) were 473806,
80464, 63708, 62977, 85187, 129931, 1801 and 20 respectively.  Using
individual timesteps Gasoline was running 4 times faster than an
equivalent single stepping code. The test was performed in parallel on
8 processors, and it is heartening that despite the added difficulties
of load balancing operations on subsets of the particles the factor of
4 benefit was realized.  Tree building and other fixed costs that do
not scale with the number of active particles can be substantially
reduced using tree-repair and related methods
\citeaffixed{springel}{e.g.}  which would bring the speed-up to around
5.  In the limit of uniform costs per particle independent of the
number of active particles the speed-up would approach 10.  In this
example the timestep range was limited partly due to restrictions
imposed on the SPH smoothing length.  In future work we anticipate a
wider range of timestep bins.  With a few particles consistently on a
timestep 1/256th of the major step the theoretical speed-up factor is
over 30.  In practice, overheads will always limit the achievable
speed-up.  In the limit of very large numbers of rungs the current
code would asymptotically approach a factor of 10 speedup.  With the
fixed costs of tree builds and domain decomposition made to scale with
the number of active particles (e.g. tree-repair), the asymptotic
speed up achievable with Gasoline would be 24 times.  The remaining
overheads are increasingly difficult to address.  For small runs, such
as the galaxy formation example used here, the low numbers of the
particles on the shortest timesteps make load balancing difficult.  If
more than 16 processors are used with the current code on this run
there will be idle processors for a noticeable fraction of the time.
Though the time to solution is reduced with more processors, it is an
inefficient use of computing resources.  The ongoing challenge is to
see if better load balance through more complex work division offsets
increases in communication and other parallel overheads.

\section{Tests}\label{tests}

\begin{figure}[h]
\epsfxsize=5.5in
\epsfbox{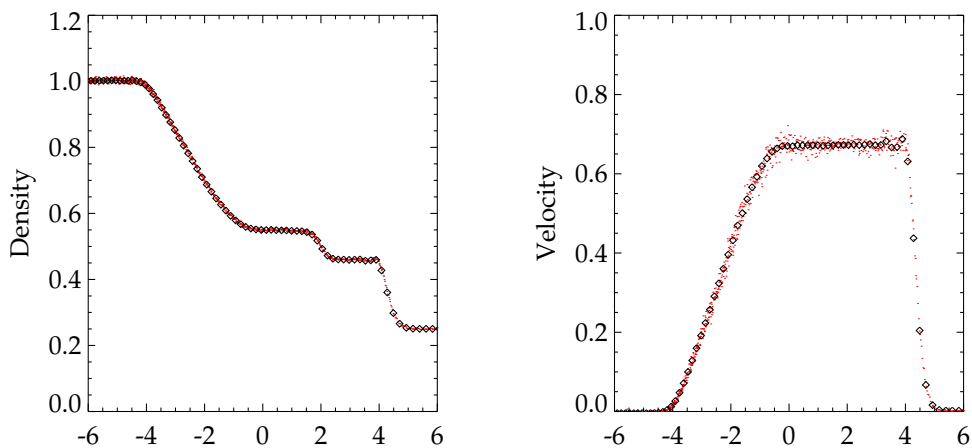}
\caption{Sod (1978) shock tube test results with Gasoline for density (left) and velocity
(right).  This a three dimensional test using glass initial conditions
similar to the conditions in a typical simulation.  The diamonds
represent averages in bins separated by the local particle spacing:
the effective fluid element width.  Discontinuities are resolved in
3-4 particle spacings which is much fewer than in the one
dimensional results shown in Hernquist \& Katz (1989).}
\label{tube}
\end{figure}

\subsection{Shocks: Spherical Adiabatic Collapse}\label{adiabaticcollapse}

There are three key points to keep in mind when evaluating the
performance of SPH on highly symmetric tests.  The first is that the
natural particle configuration is a semi-regular three dimensional
glass rather than a regular mesh.  The
second is that individual particle velocities are smoothed before they
affect the dynamics so that the low level noise in individual particle velocities
is not particularly important.  The dispersion in individual velocities is
related to continuous settling of the irregular particle distribution.
This is particularly evident after large density changes.  Thirdly, the
SPH density is a smoothed estimate.  Any sharp step in the number
density of particles translates into a density estimate that is smooth
on the scale of $\sim 2-3$ particle separations.  When relaxed irregular
particle distributions are used, SPH resolves density discontinuities
close to this limit.  As a Lagrangian method SPH can also resolve
contact discontinuities just as tightly without the advective
spreading of some Eulerian methods.

We have performed standard \citeasnoun{sod} shock tube tests used for the
original TreeSPH \cite{hk}.  We find the best results with the
pairwise viscosity of equation~\ref{artifvisc} which is marginally better
than the bulk viscosity formulation for this test .  The one
dimensional tests often shown do not properly represent the ability of
SPH to model shocks on more realistic problems.  The results of
figure~\ref{tube} demonstrate that SPH can resolve discontinuities in
a shock tube very well when the problem is setup to be comparable to the
environment in a typical three dimensional simulation.

\begin{figure}[h]
\epsfxsize=5.5in
\epsfbox{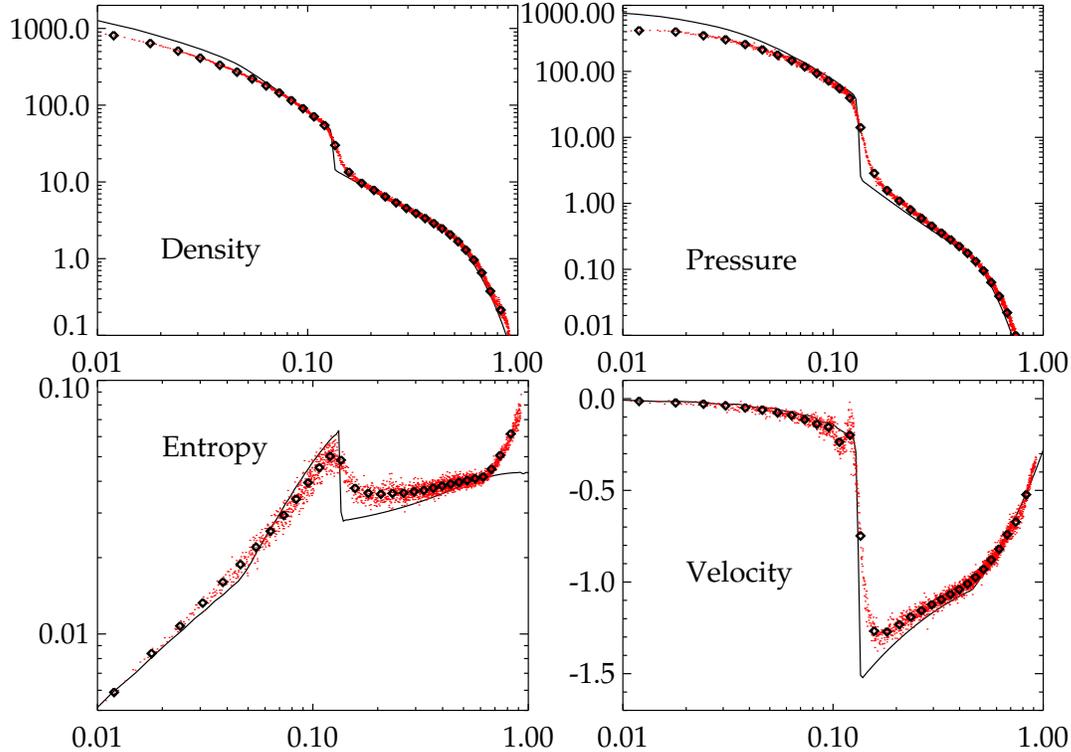}
\caption{Adiabatic collapse test from Evrard (1988) with 28000
particles, shown at time $t=0.8$ ($t=0.88$ in Hernquist \& Katz,
1989).  The results shown as diamonds are binned at the particle
spacing with actual particle values shown as points.  The solid line
is a high resolution 1D PPM solution provided by Steinmetz.}
\label{adcollfig}
\end{figure}

The shocks of the previous example have a fairly low Mach number
compared to astrophysical shocks found in collapse problems.
\cite{evrard} first introduced a spherical adiabatic collapse as a
test of gasdynamics with gravity.  This test is nearly equivalent to
the self-similar collapse of \citeasnoun{nw93} and has comparable
shock strengths.  We compare Gasoline results on this problem with
very high resolution 1D Lagrangian mesh solutions in
figure~\ref{adcollfig}.  We used a three-dimensional glass initial
condition.  The solution is excellent with 2 particle spacings
required to model the shock.  The deviation at the inner radii is a
combination of the minimum smoothing length (0.01) and earlier slight
over-production of entropy at a less resolved stage of the collapse.
The pre-shock entropy generation (bottom left panel of
Fig.~\ref{adcollfig} occurs in any strongly collapsing flow and is
present for both the pairwise (equation~\ref{artifvisc}) and
divergence based artificial viscosity formulations.  The post-shock
entropy values are correct.

\subsection{Rotating Isothermal Cloud Collapse}\label{rotatingclound}

The rotating isothermal cloud test examines the angular momentum
transport in an analogue of a typical astrophysical collapse with
cooling.  Grid methods must explicitly advect material across cell
boundaries which leads to small but systematic angular momentum
non-conservation and transport errors.  SPH conserves angular momentum
very well, limited only by the accuracy of the time integration of the
particle trajectories.  However, the SPH artificial viscosity that is
required to handle shocks has an unwanted side-effect in the form of
viscous transport away from shocks.  The magnitude of this effect
scaling with the typical particle spacing, and it can be combatted
effectively with increased resolution.  The \citeasnoun{balsara} switch
detects shearing regions so that the viscosity can be reduced where
strong compression is not also present.

We modelled the collapse of a rotating, isothermal gas cloud.  This
test is similar to a tests performed by \cite{nw93} and \cite{thacker}
except that we have simplified the problem in the manner of
\cite{ns97}.  We use a fixed \cite{nfw} (concentration $c$=11, Mass=$2
\times 10^{12} \msun$) potential without self-gravity to avoid
coarse gravitational clumping with associated angular momentum transport.  This
results in a disk with a circular velocity of 220 km/s at 10 kpc.  The
$4 \times 10^{10} \msun$, $100$ kpc gas cloud was constructed with
particle locations evolved into a uniform density glass initial
condition and set in solid body rotation ($\vec{v} = 0.407\ {\rm km/s/pc}\ 
\hat{e}_{z} \times \vec{r}$) corresponding to a rotation parameter
$\lambda \sim 0.1$. The gas was kept isothermal at $10,000$ K 
rather than using a cooling function to make
the test more general.  The corresponding sound
speed of $10$ km/s implies substantial shocking during the
collapse.

\begin{figure}
\epsfxsize=5.5in
\epsfbox{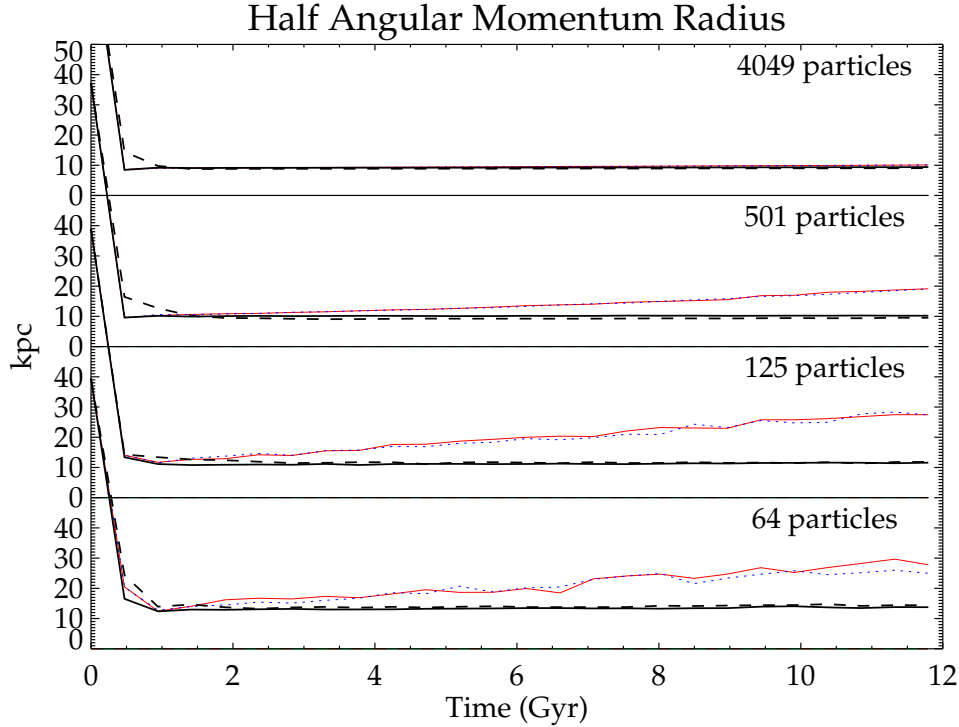}
\caption{Angular Momentum Transport as a function of resolution tested
through the collapse of an isothermal gas cloud in a fixed potential.
The methods represented are: Monaghan viscosity with the Balasara switch
(thick solid line), bulk viscosity (thick dashed line), Monaghan
viscosity (thin line) and Monaghan Viscosity without multiple
timesteps (thin dotted line). }
\label{figRotSummary}
\end{figure}

We ran simulations using 64, 125, 500 and 4000 particle clouds for
$12$ Gyr (43 rotation periods at $10$ kpc).  We present results for
Monaghan viscosity, bulk viscosity (Hernquist-Katz) and our default
set-up: Monaghan viscosity with the Balsara switch.  Results showing
the effect of resolution are shown in figure~\ref{figRotSummary}.
Both Monaghan viscosity with the Balsara switch and bulk viscosity
result in minimal angular momentum transport.  The Monaghan viscosity
without the switch results in steady transport of angular momentum
outwards with a corresponding mass inflow.  The Monaghan viscosity
case was run with and without multiple timesteps.  With multiple
timesteps the particle momentum exchanges are no longer explicitly
synchronized and the integrator is no longer perfectly symplectic.
Despite this the evolution is very similar, particularly at high
resolution.  The resolution dependence of the artificial viscosity is
immediately apparent as the viscous transport drastically falls with
increasing particle numbers.  It is worth noting that in hierarchical
collapse processes the first collapses always involve small
particle numbers.  Our results are consistent with the results of
\cite{ns97}.  In that paper self-gravity of the gas was also included,
providing an additional mechanism to transport angular momentum due to
mass clumping.  As a result, their disks show a gradual transport of
angular momentum even with the Balsara switch; however, this transport was
readily reduced with increased particle numbers.

\begin{figure}
\epsfxsize=5.5in
\epsfbox{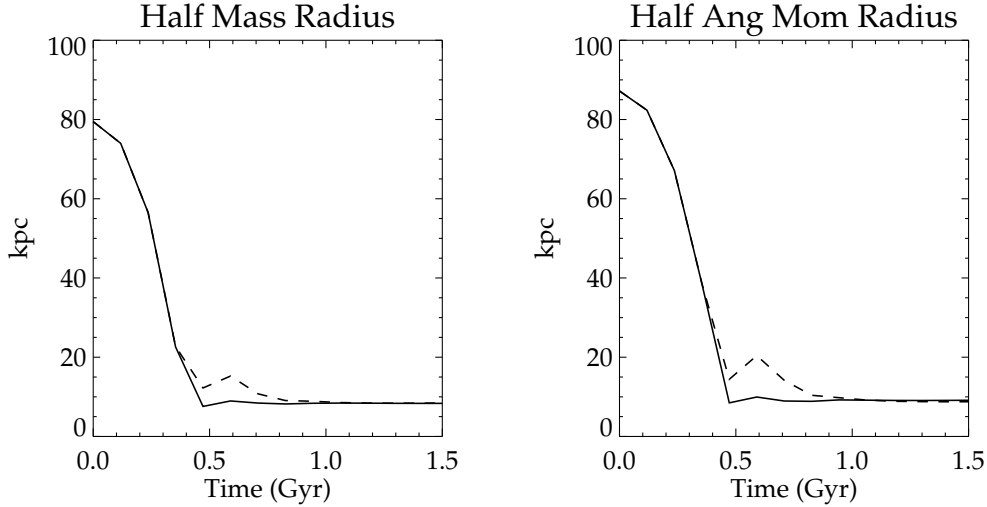}
\caption{Angular Momentum Transport test: The initial collapse phase
for the 4000 particle collapse shown in figure~\ref{figRotSummary}.
The methods represented are: Monaghan viscosity with the Balasara
switch (thick solid line) and bulk viscosity (thick dashed line).
Note the post-collapse bounce for the bulk viscosity.}
\label{figRotZoom}
\end{figure}

In figure~\ref{figRotZoom} we focus on the bounce that
occurs during the collapse using bulk viscosity.  Bulk viscosity
converges very slowly towards the correct solution and
displays unwanted numerical behaviours even at high particle numbers.
In cosmological adiabatic tests it tends to generate widespread
heating during the early (pre-shock) phases of collapse.  Thus we
favour Monaghan artificial viscosity with the Balsara switch.

\subsection{Cluster Comparison}\label{cluscomp}

\begin{figure}
\epsfxsize=5.5in
\epsfbox{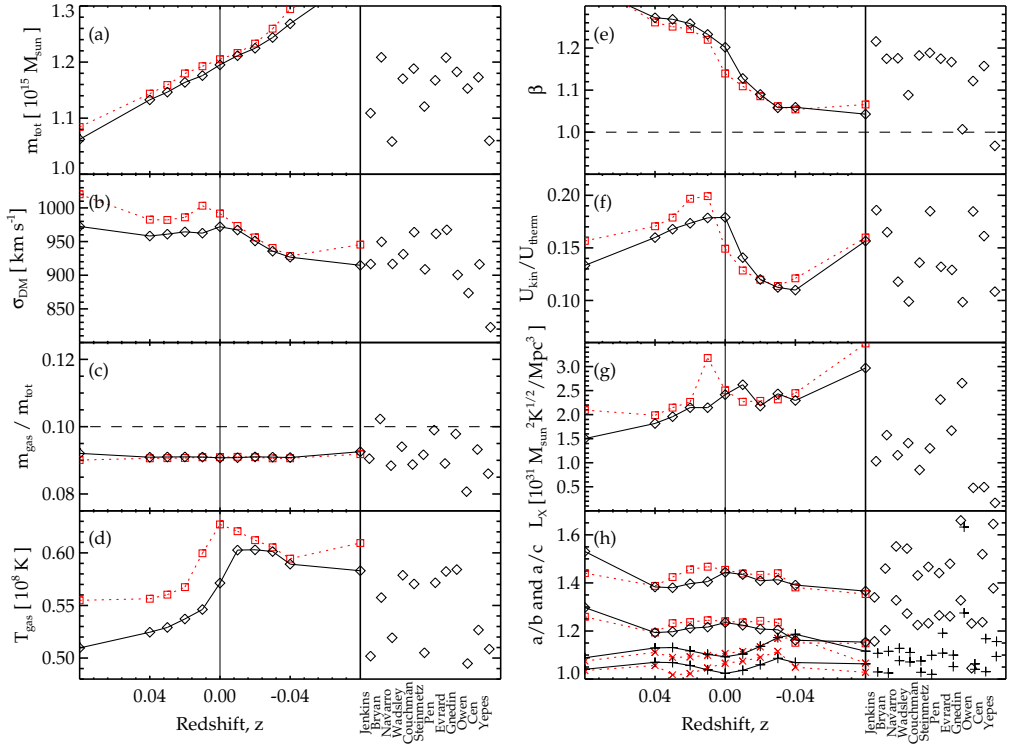}
\caption{Bulk Properties for the cluster comparison.  All quantities
are computed within the virial radius (2.74 Mpc for the gasoline
runs).  From top to bottom, the left column gives the values of: (a)
the total cluster mass; (b) the one-dimensional velocity dispersion of
the dark matter; (c) the gas mass fraction; (d) the mass-weighted gas
temperature. Also from top to bottom, the right column gives the
values of: (e) $\beta = mp/3kT$; (f) the ratio of bulk kinetic to thermal
energy in the gas; (g) the X-ray luminosity; (h) the axial ratios of
the dark matter (diamonds) and gas (crosses) distributions. Gasoline
results at $64^3$ (solid) and $128^3$ (dashed) resolution are shown
over a limit redshift range near $z=0$, illustrating issues with timing
discussed in the text.  The original paper simulation results at $z=0$
are shown on the right of each panel (in order of decreasing
resolution, left to right).  }

\label{figcluscomp}
\end{figure}

The \citeasnoun{clus} cluster comparison involved simulating the formation
of an extremely massive galaxy cluster with adiabatic gas (no
cooling).  Profiles and bulk properties of the cluster were compared
for many widely used codes.  We ran this cluster with Gasoline at two
resolutions: $2 \times 64^3$ and $2 \times 128^3$.  These were the two resolutions
which other SPH codes used in the original paper.  Profiles of the
simulated cluster have previously been shown in \citeasnoun{borgani2002}.
As shown in figure~\ref{figcluscomp}, the Gasoline cluster bulk
properties are within the range of values for the other codes at
$z=0$.  The Gasoline X-ray luminosity is near the high end; however, the
results are consistent with those of the other codes.

The large variation in the bulk properties among codes was a notable
result of the original comparison paper.  We investigated the source
of this variation and found that a large merger was occurring at $z=0$
with a strong shock at the cluster centre.  The timing of this merger
is significantly changed from code-to-code or even within the same
code at different resolutions as noted in the
original paper.  Running the cluster with Gasoline with $2 \times 64^3$
(solid) instead $2 \times 128^3$ (dotted) particles changes the timing
noticeably.  The different resolutions have modified initial waves
which affect overall timing and levels of substructure.  These
differences are highlighted by the merger event near $z=0$.  As shown
in the figure, the bulk properties are changing very rapidly in a
short time.  This appears to explain a large part of the code-to-code
variation in quantities such as mean gas temperature and the kinetic
to thermal energy ratio.  Timing issues are likely to be
a feature of future cosmological code comparisons.

\section{Applications and discussion}\label{discussion}

The authors have applied Gasoline to provide new insight into clusters
of galaxies, dwarf galaxies, galaxy formation, gas-giant planet
formation and large scale structure.  These applications are a
testament to the robustness, flexibility and performance of the code.

\citeasnoun{mqws2002} used Gasoline at high resolution to show convincingly
that giant planets can form in marginally unstable disks around
protostars in just a few orbital periods.  These simulation used 1
million gas particles and were integrated for around thousand years
until dense proto-planets formed.  Current work focusses on improving
the treatment of the heating and cooling processes in the disk.

Borgani \etal (2001,2002) used Gasoline simulations of galaxy clusters
in the standard $\Lambda$CDM cosmology to demonstrate that extreme
levels of heating in excess of current estimates from Supernovae
associated with the current stellar content of the universe are
required to make cluster X-ray properties match observations.  The
simulations did not use gas cooling, and this is a next step.  

\citeasnoun{governato2002} used Gasoline with star formation algorithms
to produce a realistic disk galaxy in the standard $\Lambda$CDM cosmology.
Current work aims to improve the resolution and the treatment
of the star formation processes.

Mayer \etal (2001a,b) simulated the tidal disruption of dwarf galaxies
around typical large spiral galaxy hosts (analogues of the Milky Way),
showing the expected morphology of the gas and stellar
components.

\citeasnoun{bond2002} used a 270 million particle cosmological Gasoline
simulation to provide detailed estimates of the Sunyaev-Zel'dovich
effect on the cosmic microwave background at small angular scales.
This simulated 400 Mpc cube contains thousands of bright X-ray
clusters and provides a statistical sample that is currently being
matched to observational catalogues \cite{wadsley512}.

A common theme in these simulations is large dynamic ranges in space
and time.  These represent the cutting edge for gasdynamical
simulations of self-gravitating astrophysical systems, and were
enabled by the combination of the key features of Gasoline described
above.  The first feature is simply one of software engineering: a
modular framework had been constructed in order to ensure
that Pkdgrav could both scale well, but also be portable to many
parallel architectures.  Using this framework, it was relatively
straightforward to include the necessary physics for gasdynamics.  Of
course, the fast, adaptable, parallel gravity solver itself is also a
necessary ingredient.  Thirdly, the gasdynamical implementation is
state-of-the-art and tested on a number of standard problems.  Lastly,
significant effort has gone into making the code adaptable to a large
range in timescales.  Increasing the efficiency with which such
problems can be handled will continue be a area of development with
contributions from Computer Science and Applied Mathematics.

\section{Acknowledgments}

The authors would like to thank Carlos Frenk for providing the cluster
comparison results and Matthias Steinmetz for providing high
resolution results for the spherical collapse.
Development of Pkdgrav and Gasoline was supported by grants from the
NASA HPCC/ESS program.

\end{document}